\documentclass[%
mathleft,%
]{an}
\usepackage{graphicx}
\usepackage{times}
\begin{document}

\Pagespan{1}{}
\Yearpublication{2012}%
\Yearsubmission{2012}%
\Month{0}%
\Volume{0}%
\Issue{0}%

\title{Non-thermal processes in coronae and beyond}

\author{K.\,Poppenhaeger\inst{1}\fnmsep\thanks{Corresponding author:
  \email{kpoppenhaeger@cfa.harvard.edu}}
\and  H.\,M. G\"unther\inst{1}
\and P. Beiersdorfer\inst{2}
\and N.\,S. Brickhouse\inst{1}
\and J.\,A. Carter\inst{3}
\and H.\,S. Hudson\inst{4}
\and A. Kowalski\inst{5}
\and S. Lalitha\inst{6}
\and M. Miceli\inst{7} 
\and S.\,J. Wolk\inst{1}
}
\titlerunning{Non-thermal processes in coronae and beyond}
\authorrunning{K.\,Poppenhaeger et al.}
\institute{Harvard-Smithsonian Center for Astrophysics, 60 Garden St, Cambridge, MA, USA
\and 
Lawrence Livermore National Laboratory, 7000 East Ave, Livermore, CA 94550, USA
\and 
University of Leicester, Department of Physics and Astronomy,
University Road, Leicester, LE1 7RH, UK
\and
SSL, UC Berkeley, 7 Gauss Way, Berkeley, CA, USA 
\and
University of Washington Astronomy Department, Box 351580, Seattle, WA
98195, USA
\and
Hamburger Sternwarte, Gojenbergsweg 112, 21029 Hamburg, Germany
\and
Dipartimento di Fisica, Universit\`a di Palermo, Piazza del Parlamento 1, 90134 Palermo, Italy}

\received{XXXX}
\accepted{XXXX}
\publonline{XXXX}


\abstract{
This contribution summarizes the splinter session ``Non-thermal processes in coronae and beyond'' held at the Cool Stars 17 workshop in Barcelona in 2012. It covers new developments in high energy non-thermal effects in the Earth's exosphere, solar and stellar flares, the diffuse emission in star forming regions and reviews the state and the challenges of the underlying atomic databases.
}

\maketitle

\section{Introduction}
While the majority of the emission from cool stars, stellar systems and the Sun is dominated by thermal processes,
non-thermal effects have been seen in recent years. When we say ``thermal effect'' here, we mean radiation that is emitted by a plasma where electrons and ions share the same temperature and their velocity distribution is Maxwellian. Furthermore, the ionization balance should be indistinguishable from the equilibrium state. We classify all emitting regions that deviate from these conditions as ``non-thermal''. Those effects are probably best observed in our own Sun, where observations can provide data with high spatial and temporal resolution. The prime example of a non-thermal effect are solar and stellar flares, where non-thermal processes can be seen in a variety of tracers. Flares might offer our best opportunity to detect non-thermal signatures in cool stars other than the Sun.

Other non-thermal processes of interest are, for example, charge exchange (CX) in solar system objects (and possibly also in young star forming regions and in exoplanetary atmospheres), and fluorescent emission in active stars.

A key challenge is how to identify and distinguish between different non-thermal mechanisms. The signals are often weak and atomic data for non-thermal processes are scarce and less well tested than collisional ionization calculations. Also, most models concentrate on thermal and time-constant emission. Thus, observers need to fully understand the uncertainty in the atomic models and modelers need to know inherent systematics in the observations. On the other hand, if non-thermal processes are unambiguously found, they provide a powerful diagnostic of plasma properties that often no other method can reveal.

To help with this task, this contribution summarizes the Splinter session ``Non-thermal processes in coronae and beyond'', which was hold as part of the Cool Stars 17 conference. The text is based on the presentations given by 
P.~Beiersdorfer, N.~S.~Brickhouse, J.~A.~Carter, H.~S.~Hudson, A.~Kowalski, S.~Lalitha, M.~Miceli and S.~J.~Wolk and the discussions which followed those talks. The talk slides are available at the splinter session home page \footnote{http://cxc.cfa.harvard.edu/cs17xraynonthermal/index.html}. In the following we group contributions based on the physical problem they describe. We start with CX in the solar system (Sect.~\ref{sect:CX}) and flares in Sect.~\ref{sect:flares}. At this point, we take a closer look at the atomic physics on which many of the equilibrium models are based and how inaccuracies in atomic databases can lead to signals that mimic non-thermal processes (Sect.~\ref{sect:atomicphysics}). This is discussed in particular in the context of the diffuse gas seen in X-ray emission in young star forming regions in Sect.~\ref{sect:starformingregions}. We end with a conclusion in Sect.~\ref{sect:conclusion}.

\section{Charge exchange in the solar system}
\label{sect:CX}
CX is found to occur in various places in the solar system. Comet observations yielded the first detection of X-rays from CX, and similar signatures are observed at several solar system planets due to the interaction of their exospheres with the solar wind (Dennerl et al. 2012). In the case of the Earth, detailed analyses can be conducted because X-ray satellites in eccentric orbits around Earth probe different locations where the solar wind interacts with Earth's exosphere. Carter et al.~2011 used the X-ray imaging CCDs of XMM-Newton to detect those CX signatures which manifest themselves as a background signal in otherwise astronomical source-free areas. The number of all archived observations between 2000 and 2009 which are detectably affected by temporarily-variable CX is around $3-4\%$. 
As expected, the signatures occurred more frequently when the
telescope's line of sight traversed the sub-solar region of the
magnetosheath, and when the Sun is at the maximum of its activity cycle.
CX signatures from a large coronal mass ejection in 2010 were clearly detected, with O~{\sc viii} as the dominant emission feature (Carter et al. 2010). The variability of this emission can be used to better understand how the Sun's and Earth's plasmas interact.

\section{Solar and stellar flares}
\label{sect:flares}
Flares are an ubiquitous phenomenon in the Sun and other cool stars, and involve a combination of thermal and non-thermal processes. The ''standard picture''  describes the starting point of a flare as a reconnection event of magnetic field lines in the corona. Electrons are accelerated during this restructuring and spiral along the magnetic field lines downwards into the chromosphere, producing gyro\-synchro\-tron radio emission. The electrons hit the denser material in the chromosphere, releasing hard X-ray radiation due to thick-target bremsstrahlung as well as optical emission due to the local heating of the chromosphere. The subsequent phase of the flare consists of chromospheric evaporation, i.e. the rise of heated material from the chromosphere into the corona, where it cools via soft (thermal) X-ray emission. Consistency with this standard picture is found in light curves from solar and stellar flares which show a distinct timing behavior summarized as the ''Neupert effect'' (Neupert 1968). The observed soft X-ray emission is proportional to the material evaporated into the corona, while the optical, radio and hard X-ray emission from the chromosphere is proportional to the chromospheric heating rate. Therefore the time derivative of the soft X-ray light curve should match the optical/radio/hard X-ray light curves. This behavior is often found in solar flares, where hard X-ray observations are readily available. For stellar flares, one uses optical and radio observations as proxies because no instruments with sufficient effective area and spatial resolution are available in the hard X-ray regime. 

It is an open question how prototypical these Neupert-like flares are in the case of stellar flares: Low-mass stars without a radiative core should not be able to operate a solar-like magnetic dynamo; consequently, their flaring behavior might differ from the solar case. Flares in the stars Proxima Centauri, an M dwarf close to the fully convective threshold, and AB Dor A, an active main sequence star of spectral type K1 were analyzed recently for timing behavior (Fuhrmeister et al. 2011). Neupert-like timing behavior was found in the optical and soft X-ray light curves of these flares, suggesting that the flaring behavior in these stars is similar to the solar case.

For the Sun itself, it is an interesting question how its corona and flaring behavior looks like during extraordinarily deep activity minima: The Solar Photometer in X-rays (SphinX) observed the solar corona in 2009 during a quiet, but on average not completely inactive phase of the solar magnetic activity cycle. The covered energy range was 1.3-14.9\,keV with a spectral resolution of $\sim460$\,eV. The observed spectra of the solar corona showed a strong thermal component with a temperature of $\sim 3$\,MK, but also a high-energy contribution which can be fitted similarly well - within the given spectral resolution - by a hotter thermal component or a non-thermal (power law) component originating from thick-target bremsstrahlung (Miceli et al. 2012). A second thermal component seems more likely because several temperature components are usually observed if active regions are present on the Sun, which was also the case during the SphinX measurements. The observations imply that even during very low activity phases small, unresolved flares heat the solar corona.

To understand the early nonthermal phases of flares, it is important to understand the particle acceleration. For the solar case, we see signatures of the primary accelerated particles; for highly energetic particles, we detect $\gamma$ rays from nuclear interactions, while moderately energetic particles ($\sim 10-100$ keV) can theoretically be observed via the Orrall-Zirker CX mechanism. In this framework, particles (H+ or He++) which are moving downward into the chromosphere undergo CX with the ambient hydrogen or helium atoms and produce red-shifted emission in the wings of the respective spectral lines. Hudson et al. (2012) have looked at the He~{\sc ii}~Ly$\alpha$ line at 304\AA{} with the Extreme Ultraviolet Variability Experiment (EVE) onboard SDO. They analyzed observations of several large solar flares to detect such CX signatures. However, no red-wing excess could be detected, which is quite surprising because highly energetic accelerated particles were present in these flares (observed as $\gamma$-ray emission).

It is difficult to compare individual phases of solar and stellar flare observations because the instruments used for the observations do not cover the same wavelength region with comparable spectral resolution. When the accelerated particle beams in flares hit the chromosphere, hard X-rays from thick-target bremsstrahlung emission can be observed in the Sun, but usually not in stars because suitable instruments with sufficient sensitivity are not available. In contrast, the optical/NUV emission from the locally heated chromosphere in stars is more easily accessible. Kowalski et al. (2012) analyzed optical/NUV spectra from large flares on several active M dwarfs to determine the nature of the enhanced emission. The excess white-light emission in flares consists of several components: a hot ($\sim 10000$\,K) black\-body-like component that displays absorption lines which resemble the spectrum of an A star, and a Balmer continuum component which mainly contributes at NUV wavelengths. To investigate the origin of the white-light emission, they modelled the flaring loop with a 1D radiative hydrodynamic code, RADYN, and injected a beam of nonthermal electrons at the top of the loop. The simulated spectra can qualitatively reproduce the Balmer continuum, but the simulation of the black\-body-like emission needs more detailed modelling, which is under way.

\section{Atomic physics}
\label{sect:atomicphysics}
In order to interpret astrophysical observations, we need to understand the physics that generates the observed signal. In high-energy physics, the emission is often optically thin and absorption can be described by very simple models, which greatly simplifies the equations describing the radiation transfer. The limiting factors on the accuracy of a spectroscopic model are then the conditions in the emitting gas and the atomic constants. How do we derive physical quantities like the temperature from the number of detected photons?  Usually, we measure the intensity of a line, which can be converted to a luminosity, if the distance to the source is known. We then fit plasma models to the spectrum. The standard model for this is the so-called collisional ionization equilibrium (CIE), where ions and electrons share the same kinetic temperature with a Maxwellian velocity distribution and the collisional excitation balances the radiative decay. This is implemented in different codes, such as APEC (Smith et al. 2001; Foster et al. 2012) and CHIANTI (Dere at al. 1998; Landi et al. 2012).

Some non-thermal processes will cause a deviation from the equilibrium ionization balance. This can be electron or ion beams where the velocity is non-Max\-well\-ian, e.g. after a magnetic reconnection where particles are magnetically accelerated or a change in the high-energy tail of the distribution caused by Fermi-acceleration in shock fronts.  Calculations of this kind require accurate knowledge of the collisional ionization and recombination (radiative, dielectronic, 2-photon) rates. Bryans et al. (2006) compare calculations and laboratory work and present revised ionization equilibria with differ up to 60\% from previous works for the peak fractional abundance.
The tails of the fractional abundances for energies away from the peak formation require larger adjustments. Unfortunately, non-equilibrium effects are seen particularly by the formation of lines at temperatures away from the peak formation temperature.
Additionally, it takes some time until Coulomb collisions equilibrate the temperatures of electrons and ions separately and later between electron and ions. After rapid heating and cooling processes, e.g. after passing of a shock wave, the gas might not be in ionization equilibrium and the measured temperature thus does not reflect the kinetic state of the plasma.

The second major source of uncertainties comes from the atomic coefficients for the collisional excitation and radiative decay. In CIE the density is assumed to be low, so most ions are in the ground state. Level population by other processes e.g. florescence or meta-stable levels thus provide a diagnostic of non-CIE conditions.
The uncertainties on collisional excitation and radiative decay are hard to quantify, but typical uncertainties can be around 25\% and in some cases much larger than this. Also, even state-of-the art atomic models might miss processes that become important for certain density and temperature conditions, particularly in non-equilibrium. Some models do not describe recombination to higher levels or simply miss weaker lines in the line lists.

Beiersdorfer \& Lepson (2012) show laboratory examples for Fe emission lines in the extreme UV. The spectra are taken on the EBIT-II electron beam ion trap at the Lawrence Livermore National Laboratory. All strong features agree well with predictions based on the CHIANTI 7.0 database and can be identified with transitions of Fe in the charge states Fe~{\sc viii} to Fe~{\sc xiv}, but over the wavelength range searched 20 weaker features remain to be explained. Some of them can be matched up with theoretically predicted iron lines. 
The situation changes when looking at spectra recorded on the NSTX plasma device, a ``fat'' tokamak at the Princeton Plasma Physics Laboratory. 
While Fe M-shell transitions are important below 250~eV, they burn out at higher temperatures. They are seen in spectra at 750~eV again, when neutral atoms enter NSTX and the gas is ionizing. The strongest differences between the spectra with $T_e = 250$~eV and 750~eV are seen in known Fe~{\sc xi} lines, which makes it suggestive to identify (some of) the new lines with Fe~{\sc xi} as well. These lines may be formed by inner shell ionization processes from a lower charge state of Fe, which may not even be in the ground state and thus could be useful markers for a hot, dense ionizing plasma.

Laboratory work like this is required to identify the transitions and verify atomic models. Particularly for instruments with low spectral resolution where many lines of different ionization stages or even different ions contribute to the observed signal (e.g. in SDO) line lists need to include all contributing lines in each filter bandpass.

It can be tempting to interpret observed spectral lines, that are not explained with a modern CIE model, as signposts of non-equilibrium physics. Any unexplained feature could be due to non-thermal effects, but, particularly if the signal is weak or the energy of the feature uncertain, a deviation between model and observations might as well be caused by inaccuracies in the CIE model.

\section{Star forming regions}
\label{sect:starformingregions}
An example for the problems discussed in Sect.~\ref{sect:atomicphysics} is the interpretation of the diffuse emission in star forming regions. Even after removing all point sources and accounting for faint undetected sources by fitting an appropriate background model several star forming regions show diffuse emission. The most extensive study comes from the Carina project (Townsley et al 2011a, 2011b), but the emission was seen in other clusters before (e.g. RCW~38: Wolk et al. 2002; M17: Townsley et al. 2003 and Broos et al. 2007; a region in the extended ONC: G\"udel et al. 2008). It is unknown what powers this extended emission. Candidates are supernovae or winds from massive stars in the center of these clusters. Curiously, all surveyed regions so far look similar.
 
Usually, the spectra are fit with one or two CIE models or a powerlaw. In Carina, the higher signal allows Townsley et al. (2011a) to fit a model of six components (three non-equilibrium plasmas and three background components accounting for unresolved stars, background galaxies and the galactic ridge). Still, they see differences between the model and the observations which could be explained by additional line-like features. Barring better alternatives, they present an interpretation suggesting that these additional lines are signatures of CX in interfaces between hot plasma and the surrounding molecular cloud.

However, the interpretation is complicated by the fact that all these observations have to be done in imaging mode (mostly with \emph{Chandra}/ACIS-I), where --compared to X-ray gratings-- the spectral resolution is much lower and a spatially and temporally varying detector background introduces additional problems. The talk of S.~Wolk showed spectra of NGC~281 which can be fitted by a CIE component and a non-thermal powerlaw, but can be described equally well with a CIE model, if the abundances of Ne, O and Fe (the most prominent emitters in the soft X-rays) are adjusted to the values typically found in active coronae.

The energy of the extra features in star forming regions is not well-determined. We have no prediction of the properties in the emitting region, thus we cannot pre-select which CX lines to look for. Many candidates exist in the X-ray range and, given the energy resolution of the CCD detectors, an observed feature at any energy will match one or more CX lines. The features are statistically significant, but are typically much less luminous than the plasma models. It is noteworthy that the detected features in Carina all have energies close to well-known strong lines in CIE models. If the atomic constants for these lines are not accurate, then this will cause deviations between the model and the data, that look as if extra lines were present. 

\section{Conclusion}
\label{sect:conclusion}
In the previous sections we discussed non-thermal emission from the Sun, from the Earth's exosphere, from extra-solar flares and from diffuse gas in star forming regions. In the Sun and our solar system the case for non-thermal processes is strong, thus we should expect them to happen in other stars as well and in fact radio observations of gyro-synchrotron emission --although not discussed in this splinter session-- show that a population of non-thermal electrons exists in some targets (G\"udel 2002). 

However, in most cases presented in the splinter session the existence of non-thermal effects in cool stars other than the Sun has not been proven. That does not mean that it is not present, it only means that, given the quality of the data, it is not \emph{required}. When many different models can explain the observed data, we usually chose the simplest model that is sufficient. If an X-ray spectrum can be described by two thermal components, then this model is typically preferred over a model of a thermal component plus a non-thermal powerlaw or plus a handful of CX lines.

So, what instrumentation would be needed to solve these issues? To investigate the solar-terrestrial connection a wide field X-ray imaging mission would be useful. For CX in solar flares an imaging instrument with high spectral resolution ($\Delta \lambda/\lambda \sim 3000$) at 304~\AA{} would enable the detection of moderately energetic particles and allow a more thorough understanding of the acceleration processes in magnetic flares. For star forming regions, we would profit from an instrument which allows a high spectral resolution for diffuse sources, i.e. a grating with a slit or a microcalorimeter with sufficient collecting area. This would help to identify any extra CX lines in the soft X-rays.

In summary, non-thermal processes in the coronae of cool stars and beyond are present, but often the signal is weak and thus they are hard to use as diagnostic tools. However, in recent years there was much progress to find more emitting sources and thus widen the horizon for more in-depth observations and simulations.



%
%
%

\end{document}